# Compressed Sensing of Scanning Transmission Electron Microscopy (STEM) on Non-Rectangular Scans

Xin Li[1,2,3,4], Ondrej Dyck[1,2], Sergei V. Kalinin[1,2]* and Stephen Jesse[1,2]*

[1]Institute for Functional Imaging of Materials, Oak Ridge National Laboratory, Oak Ridge TN 37831, USA

[2]Center for Nanophase Materials Sciences, Oak Ridge National Laboratory, Oak Ridge TN 37831, USA

[3]Department of Industrial Engineering, Florida State University, Tallahassee FL 32306, USA

[4]Department of Statistics, Florida State University, Tallahassee FL 32306, USA

This manuscript has been authored by UT-Battelle, LLC, under Contract No. DE-AC05- 00OR22725 with the U.S. Department of Energy. The United States Government retains and the publisher, by accepting the article for publication, acknowledges that the United States Government retains a non-exclusive, paid-up, irrevocable, world-wide license to publish or reproduce the published form of this manuscript, or allow others to do so, for the United States Government purposes. The Department of Energy will provide public access to these results of federally sponsored research in accordance with the DOE Public Access Plan (http://energy.gov/downloads/doe-public-access-plan)

**Abstract**

Scanning Transmission Electron Microscopy (STEM) has become the main stay for materials characterization on atomic level, with applications ranging from visualization of localized and extended defects to mapping order parameter fields. In the last several years, attention was attracted by potential of STEM to explore beam induced chemical processes and especially manipulating atomic motion, enabling atom-by-atom fabrication. These applications, as well as traditional imaging of beam sensitive materials, necessitate increasing dynamic range of STEM between imaging and manipulation modes, and increasing absolute scanning speed, that can be achieved by combining sparse sensing methods with non-rectangular scanning trajectories. Here we developed a general method for real-time reconstruction of sparsely sampled images from high-speed, non-invasive and diverse scanning pathways including spiral scan and Lissajous scan. This approach is demonstrated on both the synthetic data and the experimental STEM data on the beam sensitive material graphene. This work opens the door for comprehensive investigation and optimal design on dose efficient scanning strategies and real-time adaptive inference and control of e-beam induced atomic fabrication.

**Introduction**

Scanning transmission electron microscopy (STEM) has emerged as an extremely versatile imaging tool to analyze materials at atomic resolution(S. J. Pennycook 2017): extending from Albert Crewe's initial demonstration of single atom detection(Crewe, Wall, and Langmore 1970) to atomically resolved electron energy loss spectroscopy (EELS)(Browning, Chisholm, and Pennycook 1993; Batson 1993) and the ability of Z-contrast imaging to directly reveal structure and chemistry (Krivanek et al. 2010). The successful implementation of aberration correction not only led to record breaking resolution,(H. Rose 1974; Findlay et al. 2009; Okunishi et al. 2009; Findlay et al. 2010) but also improved contrast and revealed the potential for 3D imaging through depth sectioning enabled by high convergence angles(Albina Y Borisevich, Lupini, and Pennycook 2006; A. Y. Borisevich and Y. 2006; van Benthem et al. 2005, 2006). Another impressive achievement that has been developed in recent years is the ability of EELS to access vibrational spectroscopic signals (Krivanek et al. 2014), adding another feather in an already impressive cap of characterization tools.

The characteristic aspect enabling the EELS modality is the energy transfer from the electron beam to a material sample, including both the electronic and ionic sub-systems(Egerton 2011). One manifestation of this is electron beam damage which can be brought on by direct nuclear recoil (knock-on), beam heating, radiolysis, and induced electric fields (Jiang 2016) and is generally viewed as an effect to be avoided. Indeed, there was some question as to whether successful aberration correction would exacerbate beam damage to such an extent that it would render the advance useless (S. J. Pennycook 2017). Fortunately, it turned out not to be so. Nevertheless, various strategies have been employed to reduce electron dose such as repeated fast scans(Zhou et al. 2012) and more efficient use of the available electrons(T. J. Pennycook et al. 2015; McMullan et al. 2009, 2014). It seems clear that improvements on this front will manifest in the near future where highly efficient detectors will be coupled with sparse sampling techniques for real-time, low dose imaging.

This issue has become especially acute with the recognition that the e-beam can be used for atomic scale fabrication(Kalinin, Borisevich, and Jesse 2016; Zhao et al. 2017; Kalinin and Pennycook 2017; Jiang et al. 2017; Susi et al. 2014). Recent efforts to manipulate single atoms with the electron beam have established that such capabilities are possible, with the introduction (Dyck et al. 2017), movement (Susi, Meyer, and Kotakoski 2017; Susi et al. 2017) of atoms in a graphene lattice being demonstrated. Such abilities rely on the material or atom in question to be close to being damaged by the beam. Because the same beam is used for imaging and manipulation, one must alternate between manipulation and imaging modalities in attempts to modify and then observe the sample. This requires both increasing the dynamic range of STEM between imaging and manipulation, and increase of imaging speed, as a prerequisite to enable real-time analysis and feedback mechanisms(Jesse et al. 2015, 2018). This will act to decrease the time required for observation as well as decrease the likelihood of unintentional modification of the structure being fabricated.

Recently, the approach for sparse sensing as a pathway for low-dose STEM imaging was broadly introduced by Stevens and Browning(Andrew Stevens et al. 2014; Kovarik et al. 2016; A. Stevens et al. 2018). However, these approaches were implemented invariably for the rectangular scans of the commercial STEM systems, where the scan path is the zig-zag pattern over a rectangular area (raster scan). In other words, the electron beam is scanned from left to right, then "flies" back to the left side and scans the next row proceeding more slowly from top to bottom. A problem with this type of scan is the so-called fly-back distortion (Muller and Grazul 2001; Jones and Nellist 2013). This occurs due to hysteresis in the scan coils and is typically addressed by applying a delay between the fly-back step and the beginning of the next scan line which allows the scan coils to stabilize. For scanned images acquired over dozens of seconds, this delay may be an acceptably small fraction of the image acquisition time. However, because this delay is the same no matter how quickly the image is acquired, it may pose a significant lower bound on the speed with which acquisition can occur. The alternative scan paths introduced recently offer an improvement for scanning speed (Sang et al. 2017).

Here, we introduce a general method for real-time reconstruction of sparsely sampled images from high-speed, non-invasive and diverse scanning pathways. A critical requirement for the practical application of such a method, assuming reliable reconstruction is given, is that it must be fast enough to operate in real time (magnitudes faster than scanning duration). To ensure its applicability in this regard, it is deployed via an NVIDIA Titan X Pascal graphics processing unit (GPU) to evaluate its performance in a favorable yet realistic context (i.e. having the computational muscle of a supercomputing cluster is certainly possible but employing a local GPU-accelerated process is much more accessible and likely). The source code used for the investigations described has been made available online. This approach is demonstrated on the beam sensitive material graphene, for both the synthetic data and the operating STEM experiments.

## Materials and Methods

Real-time Image Inpainting

We want to recover the (unknown) full STEM image, $f_0$, from sparse and noisy measurements $y$ that can be obtained from various scanning pathways including spiral scan, raster scan, random scan and Lissajous scan that are demonstrated numerically and experimentally in following sections. Besides the noisy measurements, another input is the sampling mask, which indicates the spatial locations from which we have acquired the measurements, and it can be precomputed given scanning pathway and output image size. We express this in mathematical terms as,

$$y = \Phi f_0 + \epsilon \quad (1)$$

where $\epsilon$ is the measurement noise and $\Phi$ is the masking operator, defined by

$$(\Phi y)_p \stackrel{\text{def}}{=} \begin{cases} y_p & \text{if } p \in M \\ 0 & \text{otherwise} \end{cases} \quad (2)$$

where $M$ is a binary mask indicating pixel locations sampled by scanning pathways.

We focus on using structural sparsity regularization on the redundant wavelet frame (Mallat 2009) to recover an image from the sparse measurements. The reconstructed STEM image $f = Wx$ is estimated by optimizing the coefficients $x$ in the wavelet basis, where $W$ is the wavelet synthesis operator. To achieve this, we solve the $L_1$ sparsity regularized image inpainting problem via the iterative thresholding algorithm:(Daubechies, Defrise, and De Mol 2004)

$$\text{argmin}_x = \frac{1}{2}\|y - \Phi W x\|^2 + \lambda\|x\|_1 \quad (3)$$

where $\lambda$ is the tuning parameter controlling the sparsity. The soft thresholding operator plays a key role in the $L_1$ minimization that shrinks the value of coefficients:

$$S_\lambda(x) = x * \max(0, 1 - \frac{\lambda}{|x|}) \quad (4)$$

We denote the soft thresholding operator in wavelet basis as:

$$S_\lambda^W(f) = \sum_m S_\lambda(\langle f, \mathbf{w_m}\rangle)\mathbf{w_m} \quad (5)$$

where $\mathbf{w_m}$ is the basis of the wavelet frame $\mathbf{W} = \{\mathbf{w_m}\}$. Specifically, we focus on the Daubechies wavelets(Daubechies 1992), a family of orthogonal wavelets.

Following derivations in the iterative thresholding algorithm (Daubechies, Defrise, and De Mol 2004; Peyré 2011), each iteration computes:

$$f^{(i+1)} = S_\lambda^W\left(\text{ProjC}(f^{(i)})\right) \quad (6)$$

where $\text{ProjC}(f)$ computes the gradient descent of the data fidelity term $\frac{1}{2}\|y - \mathbf{\Phi W x}\|^2$:

$$\text{ProjC}(f) = \mathbf{M}y + (1 - \mathbf{M})f \quad (7)$$

To get real-time image inpainting from sparse scan measurements that is instrumental for future tasks such as real-time in-situ atom motion tracking, we developed a parallel implementation for the above iterative thresholding algorithm utilizing the CUDA toolkit of the NVIDIA Corporation. It costs around 0.008 seconds to reconstruct a STEM image (synthetic and experimental) of size 256 by 256.

Sparse Spiral Scan

We design a field-programmable gate array (FPGA)-based scan system (built upon a National Instruments PXIe-1073 chassis) capable of interfacing with various microscopes. A LabView routine is developed to generate voltage waveforms with input coordinates from customized Matlab code, enabling arbitrary and dynamic beam positioning.

Although the real-time image inpainting algorithm is adapted to various scanning pathways as demonstrated in later sections, we begin by consideration of the special shape of an Archimedean spiral that avoids abrupt changes in scanning direction and has less requirements on beam control hardware, making it an efficient and hardware-friendly scanning strategy practically. Additional discussion regarding spiral scans and the associated advantages over raster scans can be found at (Sang et al. 2017). One can directly control the total scanning duration as well as electron distribution sparseness via Archimedean spiral scan whose scanning pathway is determined by Eq. (8):

$$\mathbf{x} = A\mathbf{T}\cos(\omega\mathbf{T}), \mathbf{y} = A\mathbf{T}\sin(\omega\mathbf{T}) \quad (8)$$

Given the fixed hardware data IO_rate (the rate of reading data from microscope to computer memory) $r$, the sampling time sequence $\mathbf{T}$ is determined by the total scanning duration $D$, $\mathbf{T}= [0, D]$ with a step size of $\frac{1}{r}$. In practice, we normalize the $\mathbf{T}$ to [0,1] with the step size of $\frac{1}{rD}$. In this way, one can control the physical scanning area directly by the amplitude parameter, $A$. Taking the derivative of Eq. (8), at a particular sampling time point, t, we get the scanning beam velocity with a magnitude:

$$|\vec{v}| = A\sqrt{1 + \omega^2 t^2} \approx A\omega t \quad (9)$$

and the magnitude of angular velocity is defined by:

$$\Omega = \frac{|\vec{v}|}{r} = \frac{A\omega t}{t} = A\omega \quad (10)$$

As the sampling time increases, the spiral is drawn outward beginning at the origin with the increasing velocity magnitude. This means that the center of the spiral will be more densely sampled than the outer edge. Meanwhile, given the total scanning duration, smaller frequency $\omega$ (angular velocity) will result in a shorter total scan distance, yielding spatial sparseness of electron dose distribution, and will increase the average sampling density (this will improve the signal-to-noise ratio in practice), improving evenness of electron dose distribution. Figure 1 clearly highlight sparseness and evenness introduced by decreasing frequency ω, where we scatter points on an Archimedean spiral with $\omega = 128$, 16, and 8 (Hz) respectively.

**Results**

We performed all the experiments on the beam sensitive material graphene. CVD-grown graphene samples were transferred from Cu foil to TEM grids and cleaned via a wet transfer method and baking in an ArO2 environment as described elsewhere (Dyck et al. 2018). The dopant atom was inserted *in situ* as described elsewhere(Dyck et al. 2017). The accelerating voltage was set to be 60 kV for all the HAADF scans in this paper.

    The real-time inpainting algorithm only requires the sparse and noisy measurements $y$ together with the sampling mask **M** that can be precomputed offline. Inpainting algorithm has three primary tuning parameters: total iteration number, thresholding value and the level of wavelet. For all the synthetic and experimental datasets, we fixed total iteration number at 10, the thresholding value at 2.4. For scans with amplitude *A* of 5 (v), we choose the wavelet level to be 2. For less scanning area with the smaller amplitude *A* of 1 (v), we set the level to be 3, since the atom spans more pixels than those scanned at *A* = 5 (v). We note that reconstruction results can be further improved by fine-tuning of these parameters. To ascertain the performance of the algorithm, we first performed reconstructions with artificially sampled data with Archimedean scan. We use peak-to-signal-noise ratio (PSNR) and the structural similarity index (SSIM(Wang et al. 2004)) for measuring inpainting quality. SSIM returns value of 1 for synthetic ground-truth image compared with itself. All experiments were performed on square images of width N = 256.

Synthetic Spiral Scan

Figure 2a is the synthetic ground-truth STEM image for graphene with a single 4-fold silicon dopant without any noise, whose pixel intensities were normalized to [0,1]. We added white Gaussian noise with variance $\delta^2 = 0.80$ to the Archimedean spiral scan path imposed on the ground-truth image. For Archimedean spiral, we set frequency $\omega = 8, 16, 32, 64, 128$ (Hz) with different scanning durations: 0.1, 0.2, 0.4, 0.8, 1.6 (seconds). We fixed IO_rate *r* at 2000000 Hz and the amplitude *A* at 1(v). Figure 2b are the sparse and noisy images sampled by Archimedean spiral scans on the ground-truth image. In Figure 2b, we can see the outer region has more noise (a large view in Figure 2d) although the noise variance is fixed along the Archimedean spiral scans. This is because scanning beam moves faster at outer edges, yielding less samples within per pixel location. Figure 2c shows the impressive reconstruction images based on sparse and noisy ones.

Figure 3 and Figure 4 display the SSIM and PSNR values for all sparse and noisy measurements and the corresponding reconstruction results. Interestingly, at shortest scanning duration of 0.1 seconds, reconstruction yielded higher SSIM and PSNR values with smaller frequencies for $\omega = 32, 64, 128$ (Hz). This may be caused by higher average sampling density resulted from smaller frequency as illustrated in Figure 1.

In Figure 5, we also tested the inpainting algorithm results on the standard Lena image for a range of sparseness induced on spiral scans. We can see that the inpainting algorithm could also yield impressive results on the standard image for a general visual inspection purpose, especially for sparsely scanned images with lower frequencies $\omega$=8,16 Hz. Figure 6 are PSNR and SSIM values for reconstructions on the Lena image.

Experimental Spiral Scan

We first set the amplitude of Archimedean spiral scan at A = 5 (v), with $\omega = 16, 32, 64, 128$ (Hz) and durations of 0.1, 0.2, 0.4, 0.8, 1.6, 3.2 (seconds). Figure 7a displays the raw STEM images by Archimedean spiral scans and Figure 7b shows the corresponding inpainted results. Figure 7c is the large view of raw Archimedean spiral scanned image with maxim duration of 6.4 seconds and maxim frequency considered at $\omega = 256$ Hz. Figure 7d,e are the large views of raw and inpainted images from Archimedean spiral scan with duration of 3.2 seconds and $\omega = 32$ Hz. At amplitude voltage of 5 (v), we see the reconstruction could not yield recognizable lattice structure with $\omega$ less than 16 Hz. Also, for scans finished less than 0.8 seconds, both raw and inpainted images were severely affected by noise, making it hard to see the atomic structure.

To increase temporal resolution as well as spatial sparseness, we can straightforwardly decrease physical scanning area by reducing the amplitude of Archimedean spiral scan to A = 1 (v). For this smaller scanning area, we considered $\omega = 8, 16, 32, 64, 128$ Hz and durations of 0.1, 0.2, 0.4, 0.8, 1.6 seconds. Figure 8a displays raw STEM images by Archimedean spiral scans and Figure 8b shows the corresponding inpainted results. Figure 8c,d are the large views of raw and inpainted images from Archimedean spiral scan with duration of 0.8 seconds and $\omega = 32$ Hz. We note that scan only took 0.8 seconds on the beam sensitive material graphene, meanwhile reliable reconstruction took around 0.008 seconds.

Diverse Scanning Pathways

Besides spiral scan, the Lissajous scan is another promising scanning strategy towards video-rate, high-quality scanning imaging, whose scan trajectory is generally defined as:

$$\mathbf{x} = A_x \cos(2\pi f_x \mathbf{T}), \mathbf{y} = A_y \cos(2\pi f_y \mathbf{T}) \quad (11)$$

where $A_x, A_y, f_x, f_y$ represent the amplitudes and frequencies associated with $x, y$ axes. Previous work (Tuma et al. 2012, 2013; Bazaei, Yong, and Moheimani 2012; Yong et al. 2012) detailed the Lissajous scan properties. Here we demonstrate that the scanning duration and hardware burden can be further reduced by utilizing compressive sensing real-time inpainting algorithm.

We would like to minimize the total scanning duration with low-range frequencies, meanwhile maintain the necessary scanning denseness where acceptable image recovery can be achieved. For Lissajous scan with a finite period *Ts*, the ratio of $f_x, f_y$ has been proved (Tuma et al. 2013; Bazaei, Yong, and Moheimani 2012) to be a rational number:

$$\frac{f_x}{f_y} = \frac{2N}{2N-1} \quad (12)$$

where *N* is a positive integer and the period $T_s = \frac{1}{f_x - f_y}$. Without loss of generality, we fix $f_x$ given the scanner mechanical limitations and rewrite the period:

$$T_s = \frac{1}{f_x - f_y} = \frac{1}{f_x - \frac{2N-1}{2N}f_x} = \frac{2N}{f_x} \quad (13)$$

For spatial resolution of Lissajous scan, Bazaei et al. 2012 and Tuma et al. 2013 reported the lowest resolution zone is at the scanned area center. Following the derivation in Bazaei et al 2013, the spatial sparseness of Lissjous scan *h*, defined as the altitude length of the diamond formed by the four crossing points around the image center, can be approximately calculated via:

$$h \approx \frac{\pi A_x A_y}{N\sqrt{A_x^2 + A_y^2}} \quad (14)$$

Based on Equation (13) and (14), we can see there is a compromise between the spatial sparseness and scanning period. Spatial sparseness can be reduced by increasing *N*. This will increase the likelihood for a successful inpainting recovery. Meanwhile, scan period is proportional to *N*. A unique property of Lissajous scan is the multiresolution (Tuma et al. 2012, 2013; Bazaei, Yong, and Moheimani 2012): the resolution of Lissajous scan increase dramatically during the first half of the scan period then tends to be stable, increasing slowly until the end of period. In practice, we propose a general rule for selecting Lissajous scan parameters aided by compressive sensing:

1. Given the hardware limitations, fix the maximum frequency $f_x$.
2. Find the minimum *N* that yields acceptable image inpainting results.
3. Perform the scan only during [0: $T_s/2$] period, instead of the whole period.

Figure 9 shows reconstruction results for Lissajous scans with different N. Here we set $A_x = A_y = 1$, $f_x = 32$ Hz. We consider 4 different scanning periods $T_s = 2.0, 1.6, 1.2, 0.8$ seconds. The actual scanning duration was set to be $\frac{T_s}{2}$ (1.0,0.8,0.6,0.4 seconds). Based on equation (13), we have $N = Ts * f_x/2$ = 32, 26, 19, 13 (rounded to integer). And $f_y$ can be calculated by equation (12). In Figure 9, we can see that the inpainting algorithm starts to fail when N is decreased to 13. Figure 10 are the corresponding PSNR and SSIM values.

To further demonstrate the broad capability of real-time image inpainting algorithm, Figure11 shows experimental raster scans and inpainting results. For experimental raster scans, we apply a constant voltage on one scanning direction while another direction is driven by a sine wave. We set duration to be 1.6 seconds with $\omega = 128, 64, 32, 16, 8$ (Hz).

We also considered the random scans on the synthetic ground-truth image. Define the scanning ratio, *sr*, as the number of scanned pixels over the number of total pixels. We added the white Gaussian noises with $\delta^2 = 0.02, 0.05, 0.10$ and we set *sr* = 0.1, 0.2, 0.3, 0.4, 0.5, 0.6, 0.7, 0.8, 0.9. Figure 12a,b display the sparse and reconstructed images under various sparseness and noise levels. Figure12c,d are the corresponding SSIM and PSNR values.

**Discussion**

Recently, (Reed, Park, and Masiel 2016) reported that denoising alone may work well for fully sampled low-dose imaging. Figure 13a is a regularly scanned low-dose HAADF image over a graphene sample. The image is of 256 by 256 pixel size with a pixel dwell time of 16 us. The acquisition time was no less than 1 sec (256*256*16*1e$^{-6}$). Accelerating voltage was 60 kV, same with all the other scans in the paper. For the fully scanned low-dose image such as Figure 13a, our inpainting algorithm automatically reduced to the standard wavelet denoising algorithm (for fully sampled images, scanning mask **M**=1). Figure 13b is the denoising result by our inpainting algorithm. Figure 13c is the adaptive filter denoising result (MATLAB *wiener2* function with neighborhood of size [15,15]) and Figure 13d is the median filter denoising result (MATLAB *medfilt2* function with neighborhood of size [15,15]). For sparsely sampled images (low-frequency scan) such as Figure 8c, the "denoising only" methods may not yield acceptable reconstructions.

On reducing the scanning duration, a misconception for random sub-sampling is that the total acquisition time can be proportionally reduced by reducing number of sampled pixels (assuming the pixel dwell time fixed). However, it also takes time for the beam to move between sampled pixels. In fact, finding the shortest path connecting randomly located sampling pixels is a NP-hard problem (the travelling salesman problem). Here we explore multiple hardware-friendly scan pathways with smooth and continuous input voltages, trying to reduce beam movement distance/time to cover the scanned area, at the cost of sparseness that may be relieved by inpainting algorithm.

We provided a variety of scanning settings to show the limits for the inpainting algorithm. Both the sparseness and unevenness in the sampling density will affect the success of reconstruction. For successful reconstructions, neither the frequency nor duration can be too low. For high degree of sparseness as in Fig 11k, the inpainted image shows more of the scan pattern than the sample. Increasing frequency does not necessarily reduce the negative effects on reconstruction made by decreasing the duration. For example, in Figure8, we can see the reconstruction under w=128 Hz, D=0.1 sec is worse than that under w=16 Hz, D=1.6sec. Future work could be the establishment of a rigorous and systematic mathematical optimization framework for finding "optimal" scanning pathways accounting for hardware constrains as well as sparseness distribution where inpainting algorithm works. For example, the inner region of spiral scan is more densely sampled, on the contrary, the outer area of Lissajous scan is more densely sampled. A "optimal" scanning pathway may be designed to combine advantages of spiral and Lissajous scans, with a more uniformly distributed sampling density (dose rate).

In the foreseeable future, the very same e-beam is used for both imaging mode and fabrication mode. With little hardware updates and cost, our work increases the imaging temporal resolution, enabling the real-time imaging feedback mechanism. One immediate follow-up is incorporating current setting into live data stream from microscope to perform imagining-driven, adaptive inference of e-beam induced dynamical systems, utilizing recently developed statistical learning approaches. Another uptake for the STEM community is that our work opens the door for the comprehensive investigation and optimal design on dose efficient scanning strategies. To further stimulate the progress for STEM community, we opensource the code online.

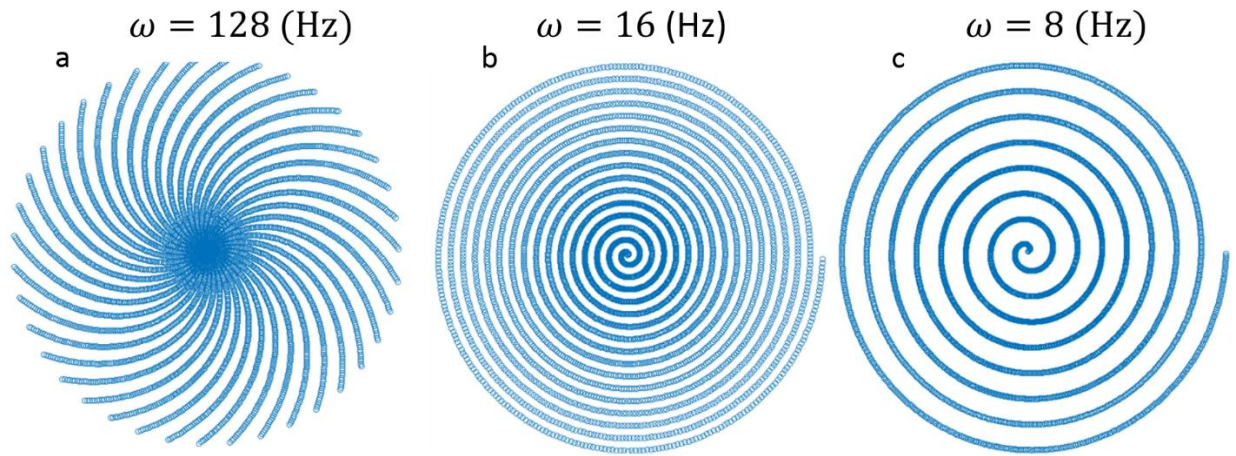

**Figure 1**: Dose distribution sparseness and evenness induced by small ω on Archimedean spiral scan trajectory: Scatter plots of (*x,y*) coordinates in scan trajectories with scanning frequencies (a) $\omega = 128$ (Hz), (b) $\omega = 16$ (Hz) and (c) $\omega = 8$ (Hz). For all the three scanning frequencies, we fixed IO_rate *r* = 200000 Hz, total scanning duration *D* = 0.05 seconds and amplitude *A* = 1 v.

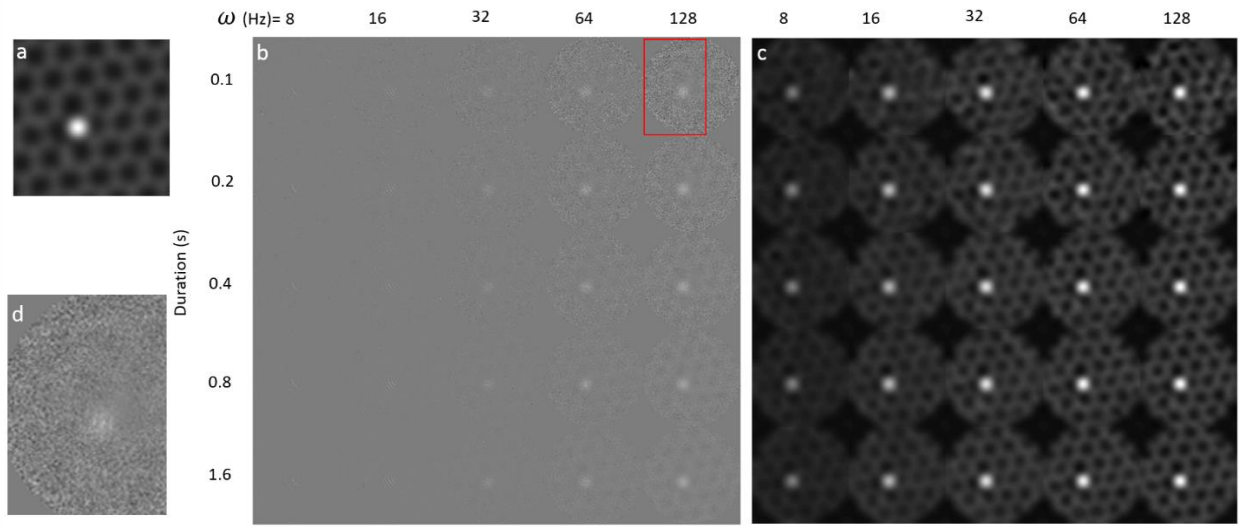

**Figure 2:** Synthetic Spiral Scan. (a) Ground-truth image. (b) Sparse and noisy Archimedean spiral scans over the ground-truth image. (c) Reconstructed images. (d) A larger view inside the marker in (b).

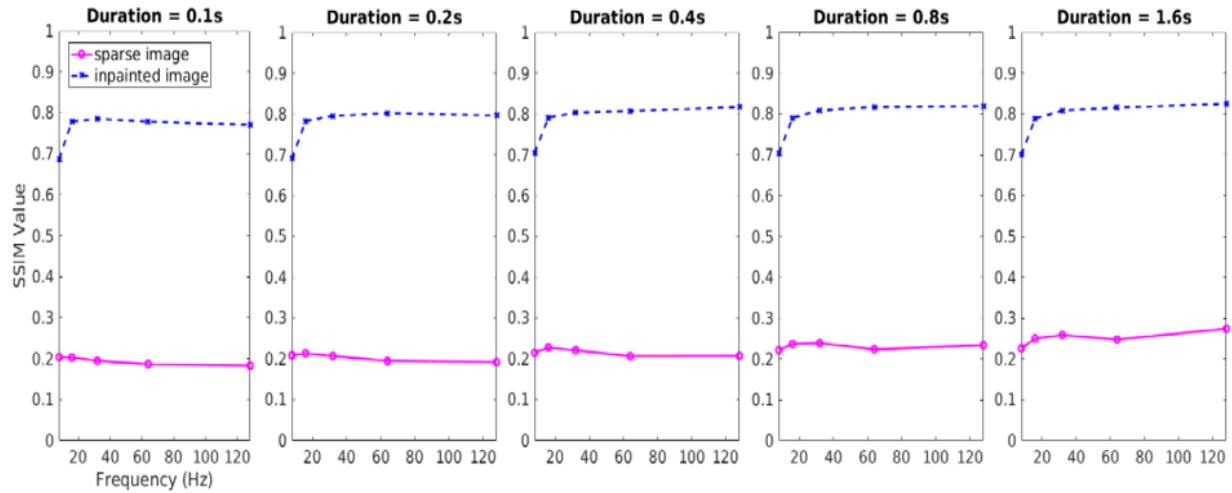

**Figure 3**: SSIM values of sparse and reconstruction images on synthetic spiral scans.

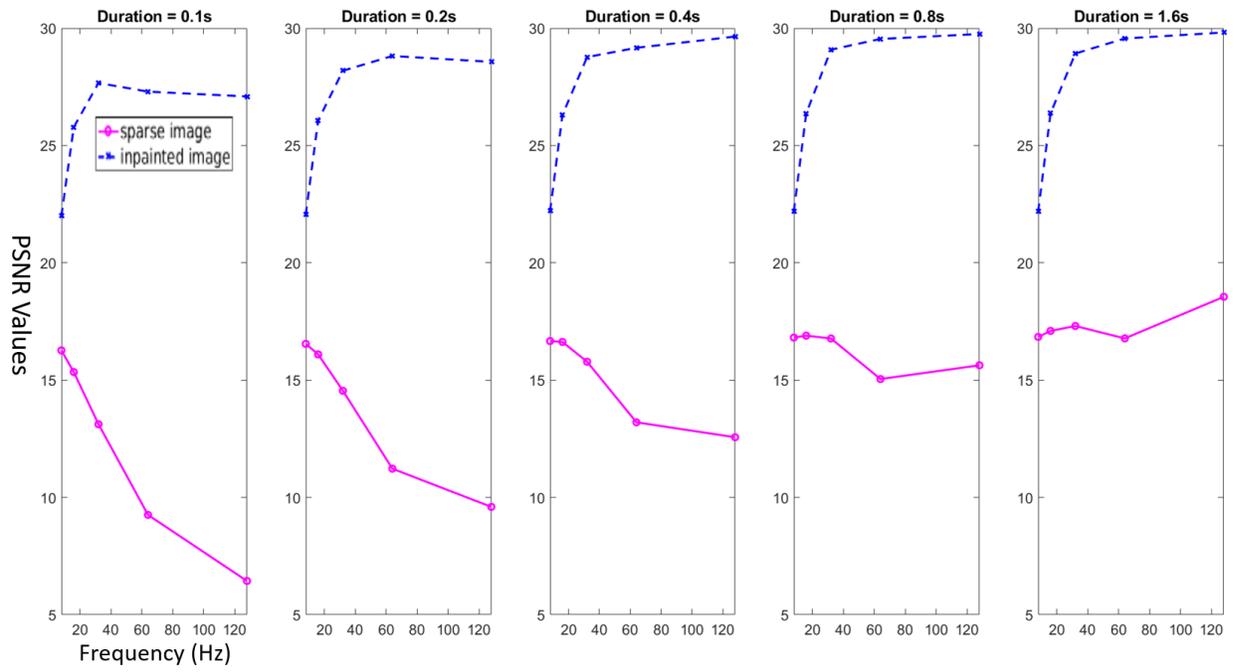

**Figure 4:** PSNR values of sparse and reconstruction images on synthetic spiral scans.

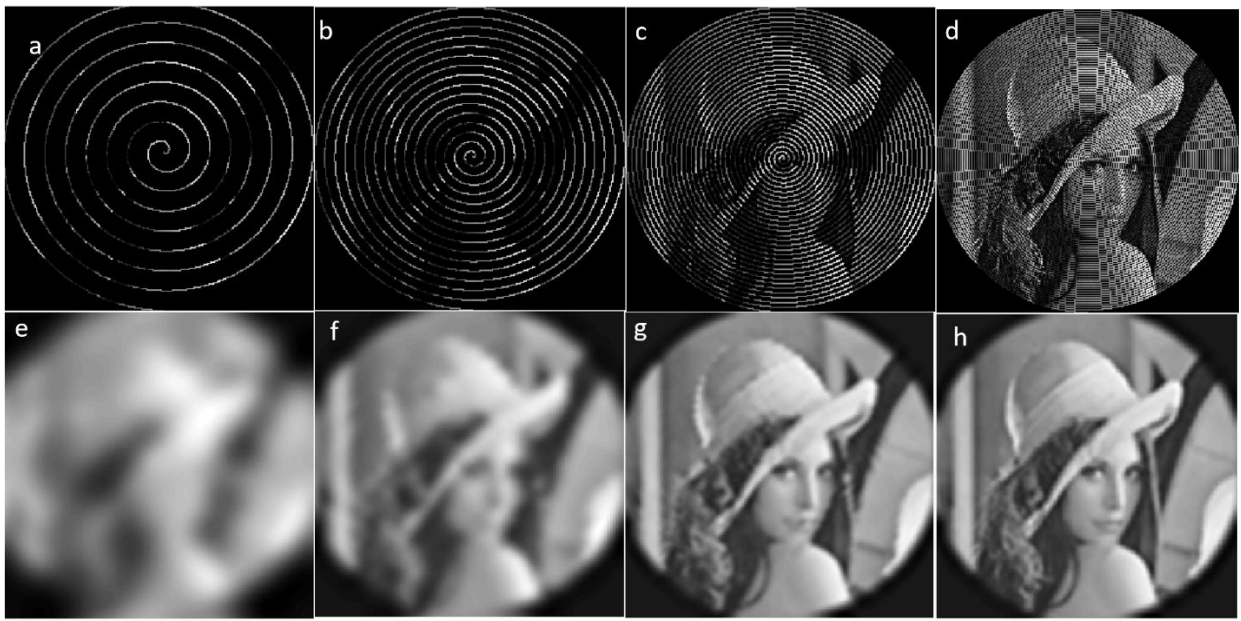

**Figure 5:** Sparse spiral scans and reconstructions on the Lena image.

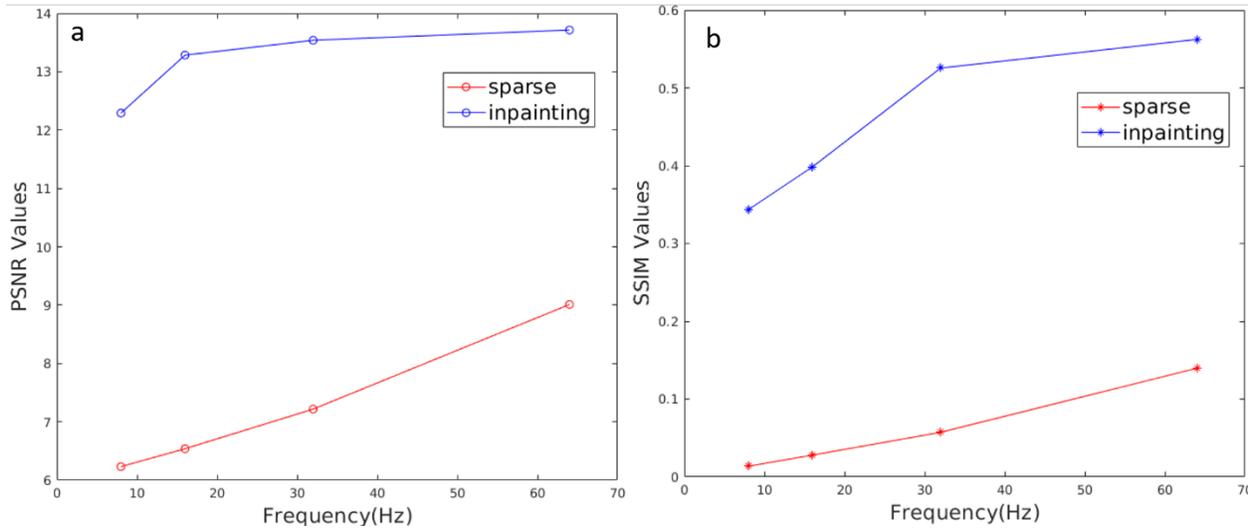

**Figure 6:** PSNR and SSIM values sparse spiral scans and reconstruction images on the Lena image.

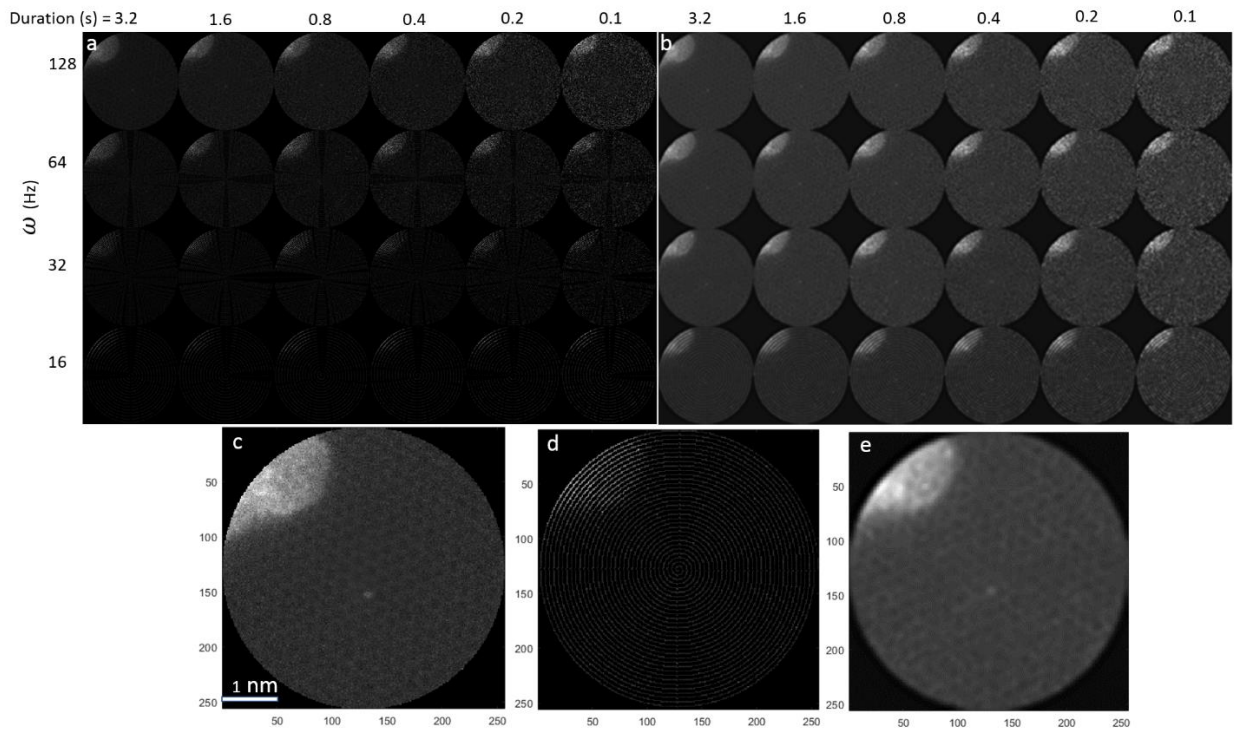

**Figure 7:** Experimental Archimedean spiral scans with voltage of *A* =5 (v). (a,b) Raw and inpainted images. (c) Raw STEM image from Archimedean spiral scan with duration of 6.4

seconds and frequency of 256 Hz. (d,e) Large views of raw and inpainted images from Archimedean spiral scan with duration of 3.2 seconds and frequency of 32 Hz.

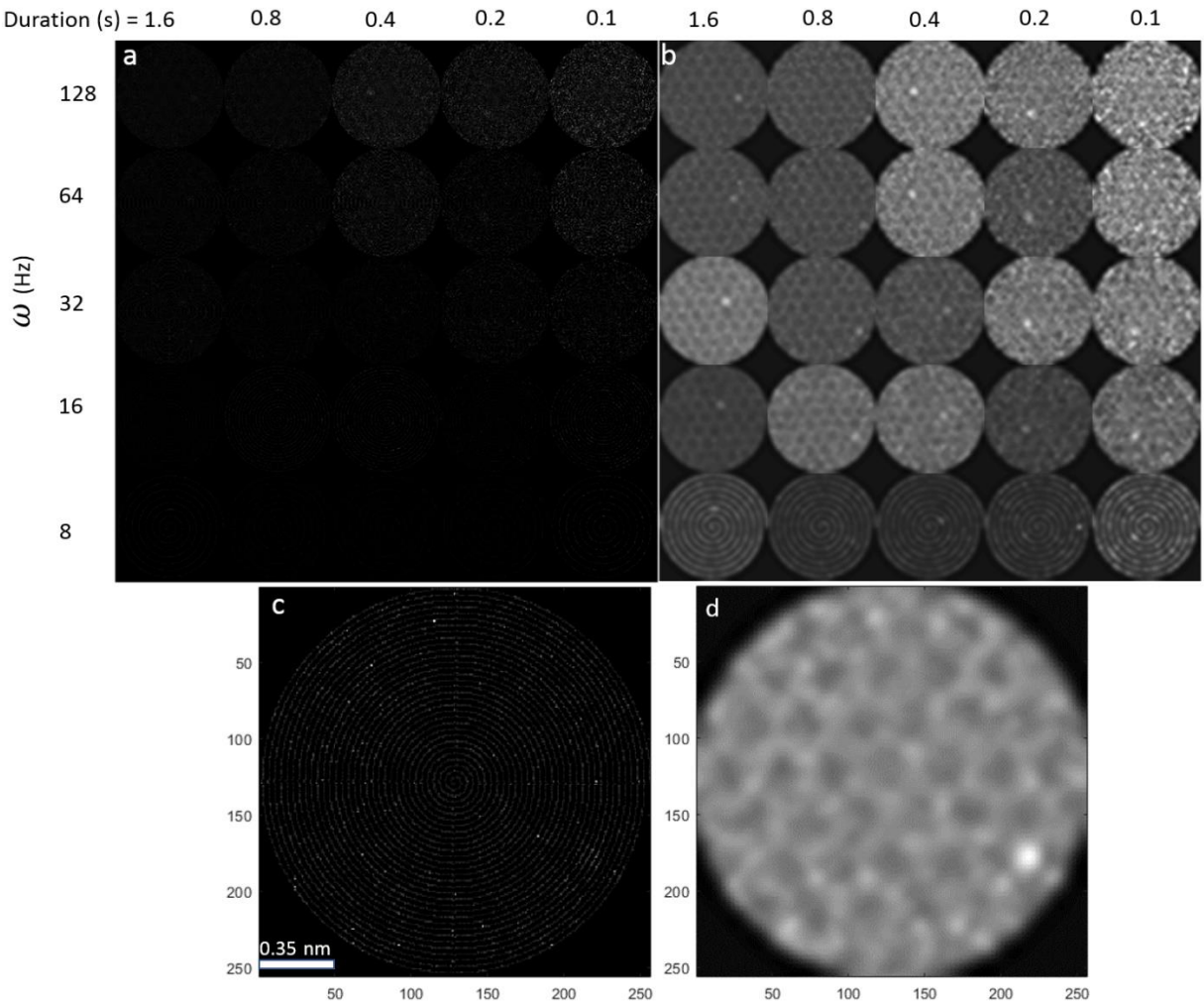

**Figure 8:** Experimental Archimedean spiral scans with voltage of $A = 1$ (v). (a,b) Raw and inpainted images. (c,d) Large views of raw and inpainted images from Archimedean spiral scan with duration of 0.8 seconds and frequency of 32 Hz.

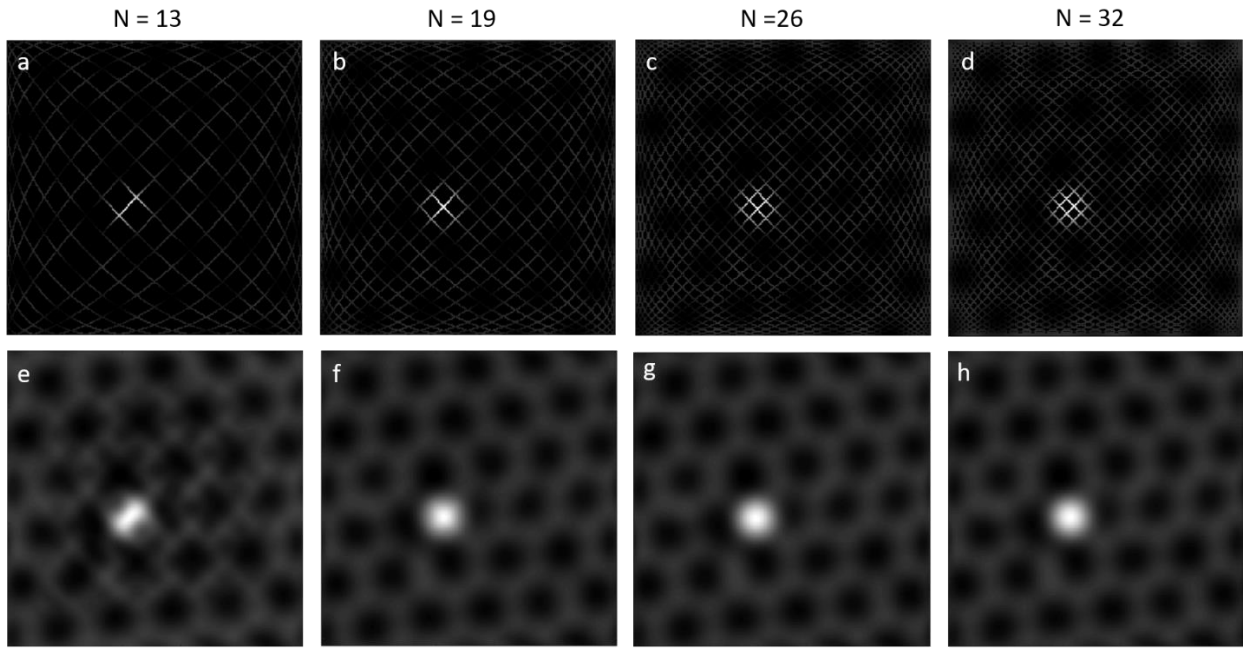

**Figure 9:** Synthetic Lissajous scans. (a-d) Sparse and noisy images. (e-h) Corresponding inpainting results.

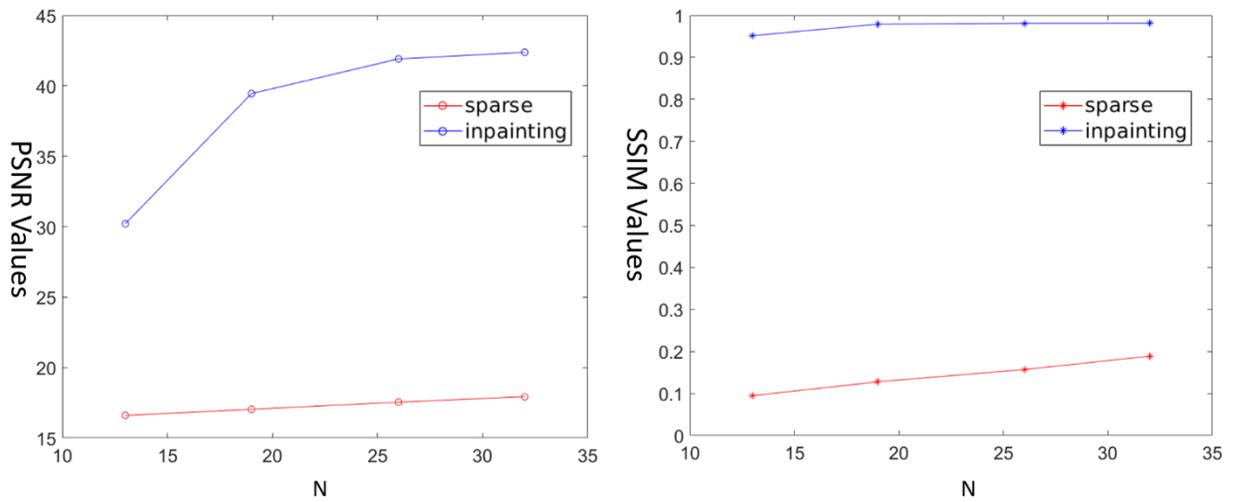

**Figure 10:** PSNR and SSIM values of sparse and reconstruction images on synthetic Lissajous scans.

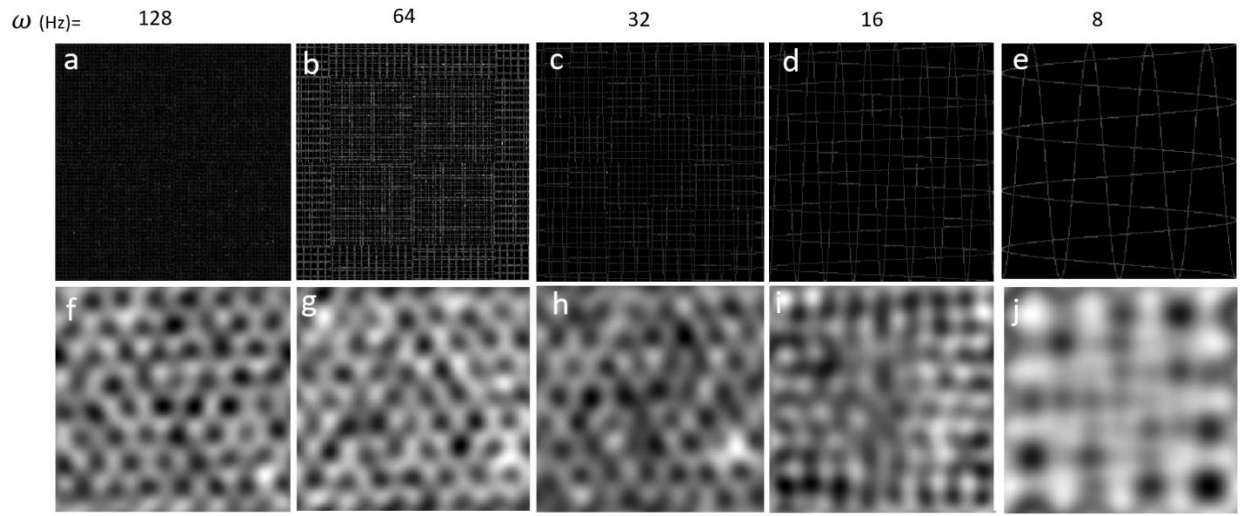

**Figure11:** Experimental raster scans. (a-e) Sparse and noisy images. (f-j) Corresponding inpainting results.

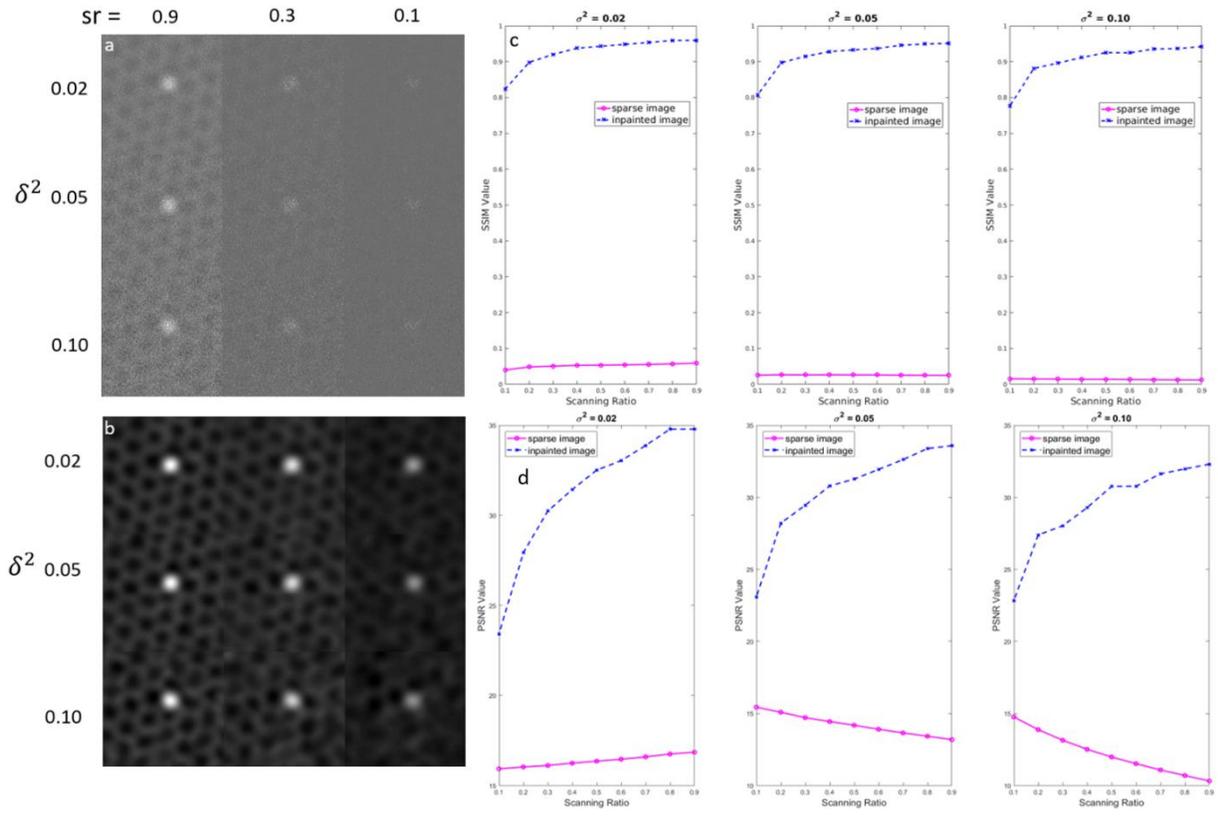

**Figure12:** Synthetic random scans. (a) Sparse and noisy images. (b) Inpainted results. (c,d) Numerical evaluations by SSIM and PSNR.

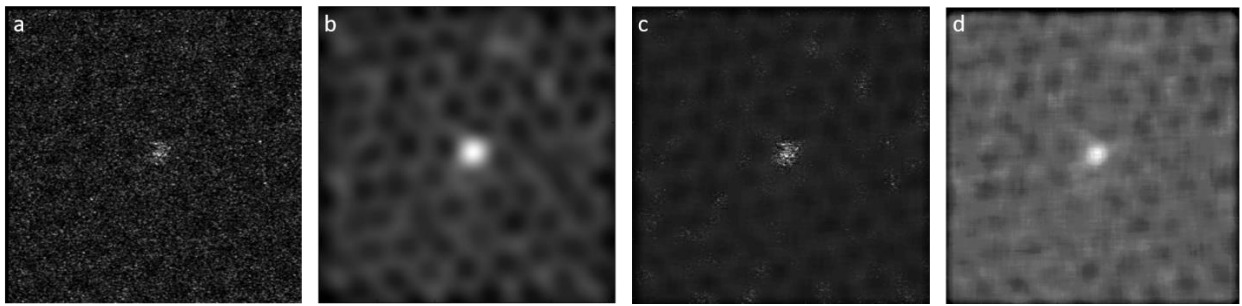

**Figure 13:** (a) A regularly scanned low-dose HAADF image over a graphene sample. The image is of 256 by 256 pixel size with a pixel dwell time of 16 us. The accelerating voltage is 60 kV. (b) Denoising results by the proposed inpainting algorithm. (c) Denoising result by the adaptive filtering. (d) Denoising result by the median filter.

**Data Availability**

The data that support the findings of this study are available from the corresponding authors upon request. The source code can be found at https://github.com/nonmin/RTSSTEM. The wavelet transform is based on the PDWT(Paleo 2016) package.

**Acknowledgements**

This research was conducted at the Center for Nanophase Materials Sciences, which is a DOE Office of Science User Facility. S.V.K and S.J. acknowledges the DOE FWP program. We gratefully acknowledge the support of NVIDIA Corporation with the donation of the Titan X Pascal GPU used for this research.